\documentclass[twocolumn]{aastex62}

\usepackage{acronym}
\usepackage{amsmath}
\usepackage{graphicx}	
\DeclareGraphicsExtensions{.pdf,.png,.jpg,.ps}
\graphicspath{{./},{./plots/}}
\acrodef{SN}{supernova}
\acrodefplural{SN}[SNe]{supernovae}
\acrodef{CCSN}{core-collapse supernova}
\acrodefplural{CCSN}[CCSNe]{core-collapse supernovae}
\acrodef{PISN}{pair-instability supernova}
\acrodefplural{PISN}[PISNe]{pair-instability supernovae}
\acrodef{O/C}{oxygen-carbon}
\acrodef{ZAMS}{zero-age main sequence}
\acrodef{MS}{main sequence}
\acrodef{CSM}{circumstellar medium}

\newcommand{\Msol}[0]{\ensuremath{\rm M_\odot}}
\newcommand\event{iPTF14hls}
\newcommand\rate{\mathcal{R}}

\definecolor{mustard}{rgb}{1.0, 0.86, 0.35}
\definecolor{cyan(process)}{rgb}{0.0, 0.72, 0.92}
\definecolor{ochre}{rgb}{0.8, 0.47, 0.13}

\newcommand\sixtySixtyMergerSMCHydrogen{9.8}

\newcommand\ninetyFiveSingleSMCHelium{6.0}

\newcommand\sixtySixtyMergerSMCHelium{19.0}

\newcommand\lowerLimitPPISN{28}
\newcommand\upperLimitPPISN{52}

\newcommand\lowerLimitMESA{40}
\newcommand\upperLimitMESA{64}

\newcommand\lowerLimitLogCCSNRate{1.3\times10^{-5}}
\newcommand\upperLimitLogCCSNRate{3.2\times10^{-3}}
\newcommand\primaryNarrow{0.01}
\newcommand\primaryWide{0.06}
\newcommand\separationShort{0.13}
\newcommand\separationLong{0.46}

\shorttitle{Pulsational PISNe following stellar mergers}
\shortauthors{Vigna-G\'omez et al.}
\begin{document}

\title{Massive Stellar Mergers as Precursors of Hydrogen-rich Pulsational Pair Instability Supernovae}

\author[0000-0003-1817-3586]{Alejandro Vigna-G\'omez}
\email{avigna@star.sr.bham.ac.uk}
\affiliation{Birmingham Institute for Gravitational Wave Astronomy, University of Birmingham, Birmingham, B15 2TT, UK}
\affiliation{Monash Centre for Astrophysics, School of Physics and Astronomy, Monash University, Clayton, Victoria 3800, Australia}
\affiliation{DARK, Niels Bohr Institute, University of Copenhagen, Blegdamsvej 17, 2100, Copenhagen, Denmark}

\author{Stephen Justham}
\affiliation{School of Astronomy \& Space Science, University of the Chinese Academy of Sciences, Beijing, China}
\affiliation{National Astronomical Observatories, Chinese Academy of Sciences, Beijing 100012, China}
\affiliation{Anton Pannekoek Institute for Astronomy, University of Amsterdam, 1090 GE Amsterdam, The Netherlands}
\affiliation{GRAPPA, University of Amsterdam, Science Park 904, 1098 XH Amsterdam, The Netherlands}
\affiliation{DARK, Niels Bohr Institute, University of Copenhagen, Blegdamsvej 17, 2100, Copenhagen, Denmark}

\author[0000-0002-6134-8946]{Ilya Mandel}
\affiliation{Monash Centre for Astrophysics, School of Physics and Astronomy, Monash University, Clayton, Victoria 3800, Australia}
\affiliation{Birmingham Institute for Gravitational Wave Astronomy, University of Birmingham, Birmingham, B15 2TT, UK}
\affiliation{DARK, Niels Bohr Institute, University of Copenhagen, Blegdamsvej 17, 2100, Copenhagen, Denmark}

\author[0000-0001-9336-2825]{Selma E. de Mink}
\affiliation{Anton Pannekoek Institute for Astronomy, University of Amsterdam, 1090 GE Amsterdam, The Netherlands}
\affiliation{GRAPPA, University of Amsterdam, Science Park 904, 1098 XH Amsterdam, The Netherlands}
\affiliation{DARK, Niels Bohr Institute, University of Copenhagen, Blegdamsvej 17, 2100, Copenhagen, Denmark}

\author[0000-0002-8338-9677]{Philipp Podsiadlowski}
\affiliation{Department of Astronomy, Oxford University, Oxford OX1 3RH, United Kingdom}
\affiliation{DARK, Niels Bohr Institute, University of Copenhagen, Blegdamsvej 17, 2100, Copenhagen, Denmark}

\begin{abstract}
Interactions between massive stars in binaries are thought to be responsible for much of the observed diversity of supernovae. As surveys probe rarer populations of events, we should expect to see supernovae arising from increasingly uncommon progenitor channels. Here we examine a scenario in which massive stars merge after they have both formed a hydrogen-exhausted core. We suggest this could produce stars which explode as \acp{PISN} with significantly more hydrogen, at a given metallicity, than in single-star models with the same pre-explosion oxygen-rich core mass. We investigate the subset of those stellar mergers which later produce pulsational \acp{PISN}, and estimate that the rate of such post-merger, hydrogen-rich pulsational \acp{PISN} could approach a few in a thousand of all core-collapse supernovae. The nature and predicted rate of such hydrogen-rich pulsational \acp{PISN} are reminiscent of the very unusual supernova \event. For plausible assumptions, PISNe from similar mergers might dominate the rate of PISNe in the local Universe.
\end{abstract}


\section{Introduction} \label{sec:intro}

The diversity of ways in which massive stars die is heavily affected by the interactions which they undergo during their lifetimes \citep{Podsiadlowski+1992}.  It is natural to consider binary-star pathways towards all outcomes of massive star evolution, as these stars are typically born in interacting binaries \citep{Sana:2012}.  Moreover, since the binary-interaction parameter space is large and multi-dimensional, ongoing \ac{SN} discoveries may reveal new diversity arising from novel binary evolution routes.  One relatively unexplored question is the influence of binarity on the diversity of pair-instability supernovae (PISNe). 

\acp{PISN} are predicted to occur in stars with sufficiently massive \ac{O/C} cores. In those cores, the temperatures become high enough for photons to produce electron-positron pairs, which results in a decrease of the radiation pressure support, and can lead to a dynamical instability. The core then contracts until the temperature is high enough to initiate explosive oxygen fusion \citep{barkat1967dynamics,rakavy1967carbon}. If nuclear burning reverses the contraction, this process may result in a single explosion as a \ac{PISN}, completely disrupting the star.  

Models of less massive stellar cores, with \ac{O/C} core masses between approximately $\lowerLimitPPISN$ and $\upperLimitPPISN~{\rm{M_{\odot}}}$\footnote{This range is found for pure helium cores in \cite{Woosley:2017}. Different assumptions lead to different ranges, e.g., \cite{chatzopoulos2012effects}.} 
find that they can avoid disruption by the first pulse of explosive burning \citep{barkat1967dynamics,Woosley:2017}. Those stars may experience multiple pair-instability eruptions, collectively called pulsational \acp{PISN}, after which there is insufficient nuclear fuel remaining to reverse the next collapse. Pulsational \acp{PISN} have been proposed as being responsible for a possible limit to the masses of black holes so far detected by gravitational wave detections \citep{LIGO-GWTC2018}, and as explanations for some very luminous \acp{SN} \citep{Woosley:2007}, but there has been no unambiguous identification of such a pair-instability driven stellar death.

Most newly-observed \acp{SN} fit within existing classes, but \event\ \citep{arcavi2017energetic} is an extraordinary exception. The inferred bolometric luminosity stayed above $10^{42}~\rm{erg~s^{-1}}$ for over 600 rest-frame days, a duration more than 6 times longer than that of a canonical \ac{SN}, and the light curve during this time displayed at least 5 peaks. This is different from the single plateaus seen in other hydrogen-rich Type II-P \acp{SN} to which \event\ is spectroscopically similar \citep{arcavi2017energetic}. The total energy radiated during those 600 days was a couple of times $10^{50}~\rm{erg}$, well above the energy inferred for any previously known Type II-P \ac{SN}. After 600 days, \event\ remained more luminous than a typical Type II-P \ac{SN} at peak luminosity. The multiple-peaked light curve is somewhat reminiscent of a pulsational \ac{PISN} \citep{barkat1967dynamics,Woosley:2017}.   However, one challenge with interpreting \event\ as a pulsational \ac{PISN} is that single stars sufficiently massive to produce luminous pulsational \acp{PISN} at metallicities similar to that of the apparent host galaxy of \event\ are typically expected to retain too little hydrogen by the time of the explosion to produce the observed hydrogen-rich \event\ \citep{arcavi2017energetic,Woosley:2018}.  

\begin{figure*}
\includegraphics[width=\textwidth]{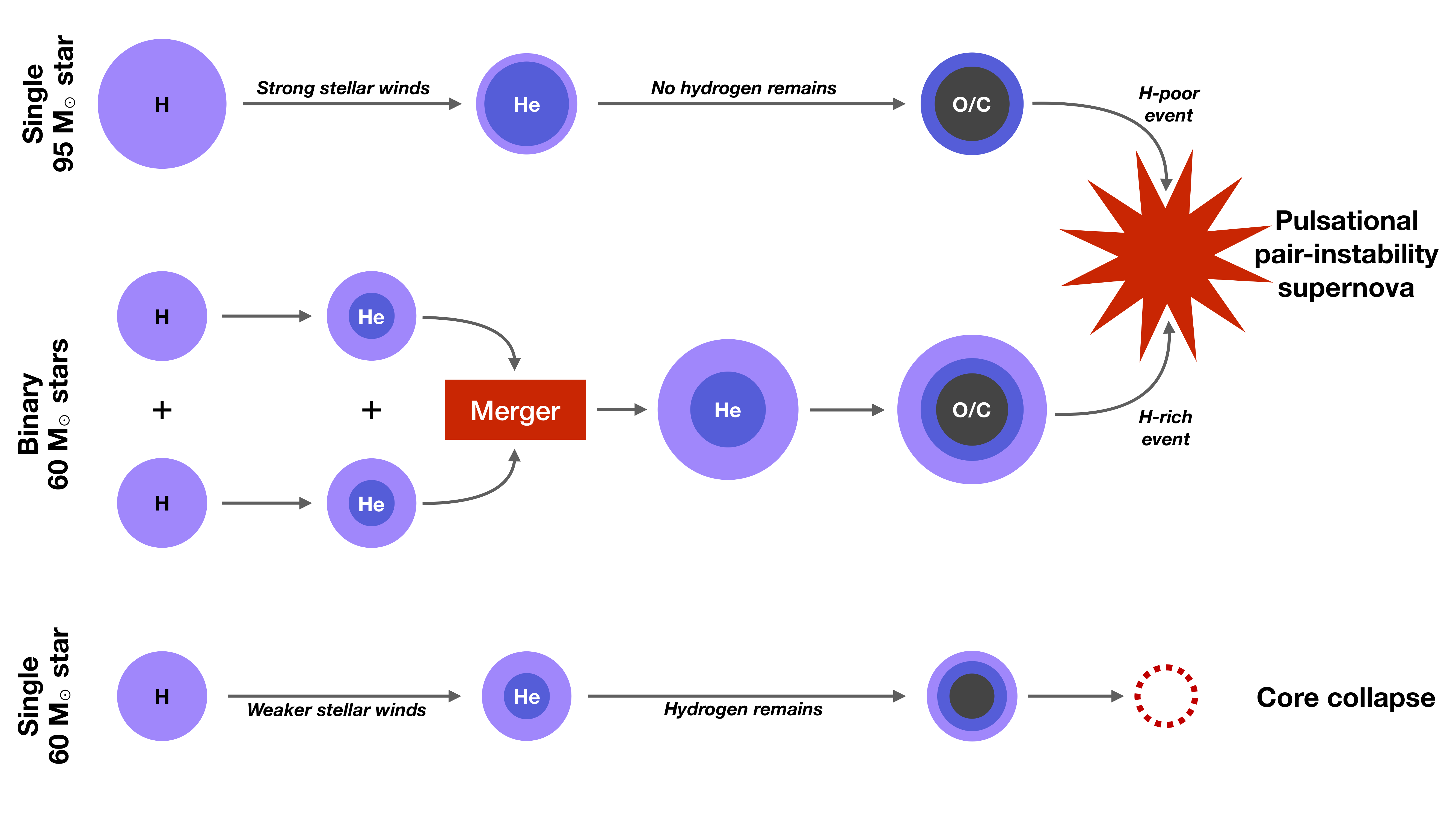}
\caption{Schematic representation of three possible outcomes of stellar evolution. Top:  single stellar evolution of a $95\ \rm{M_{\odot}}$ star leading to a canonical pulsational \ac{PISN}. Middle: merger scenario of two stars of initially $60\ \rm{M_{\odot}}$ each, as investigated in this Letter; this scenario leads to a hydrogen-rich pulsational \ac{PISN}. Bottom: single stellar evolution of non-rotating $60\ \rm{M_{\odot}}$ star, which doesn't produce a carbon-oxygen core massive enough to be a pulsational \ac{PISN}. \label{fig:cartoon}}
\end{figure*}

Here we investigate whether a class of stellar mergers could produce stars that later explode in a hydrogen-rich pulsational \ac{PISN}.  In this model, two stars in a binary system merge to form the \ac{SN} progenitor after the end of the relatively long phase of hydrogen fusion in each of their cores \citep{justham2014luminous}.   A merger during this evolutionary stage is natural, due to the expansion of the stars after the end of their core hydrogen fusion, and the pre-merger stars will have experienced significantly less fractional mass loss by winds than a single star of the same total mass and evolutionary state.  The merger creates a combined helium core sufficiently massive to lead to a pair-unstable \ac{O/C} core.    Figure \ref{fig:cartoon} displays a schematic representation of a single star leading to a pulsational \ac{PISN}, as well as this merger formation channel.   We find that this progenitor scenario could lead to a more substantial hydrogen-rich envelope at the onset of the explosion than in single-star pulsational \ac{PISN} models with otherwise identical assumptions.    We estimate that this evolutionary route could well be significant in producing pulsational \acp{PISN} in the local Universe.

\section{Massive Stellar Mergers} \label{sec:mergers}
\subsection{Astrophysical Case}
Two similarly massive stars transferring mass at an appropriate orbital separation can merge. When both stars are expanding after their \ac{MS}, mass transfer can cause the accreting star to also over-fill its Roche lobe. Subsequent loss of mass and angular momentum from one or both of the outer Lagrangian points can cause runaway shrinking of the binary orbit. For such stars, it appears likely that this situation would at least sometimes, and perhaps typically, lead to a merger (see, e.g., \citealt{Pasquali+2000};  \citealt{Podsiadlowski2010}; \citealt{justham2014luminous}).  

Cases in which unstable mass transfer occurs when two similar-mass stars have completed their \ac{MS}, but in which successful common-envelope ejection prevents a merger, have been proposed as a pathway to explain the formation of some double-neutron-star binaries \citep{brown1995doubleCore,dewi2006double} and low-mass binary systems \citep{Justham+2011}. Those cases typically require binary components with initial masses similar to within, at most, a few per cent.  However, at higher masses the probability for two stars to interact during the post MS evolution of both stars becomes larger.  This is because the duration of the \ac{MS} becomes a very shallow function of initial mass  \citep[see, e.g.,][]{Brott:2011}. Hence stars from a wider relative range in mass can simultaneously be between the end of core hydrogen fusion and the start of core oxygen fusion, ideal for mergers in our scenario. (Such systems with similar masses and appropriate orbital periods have been observed, e.g.\ R139/VFT527, for which see \S \ref{sec:discussionAndConclusions}.) 

\begin{figure*}
	\includegraphics[width=\textwidth]{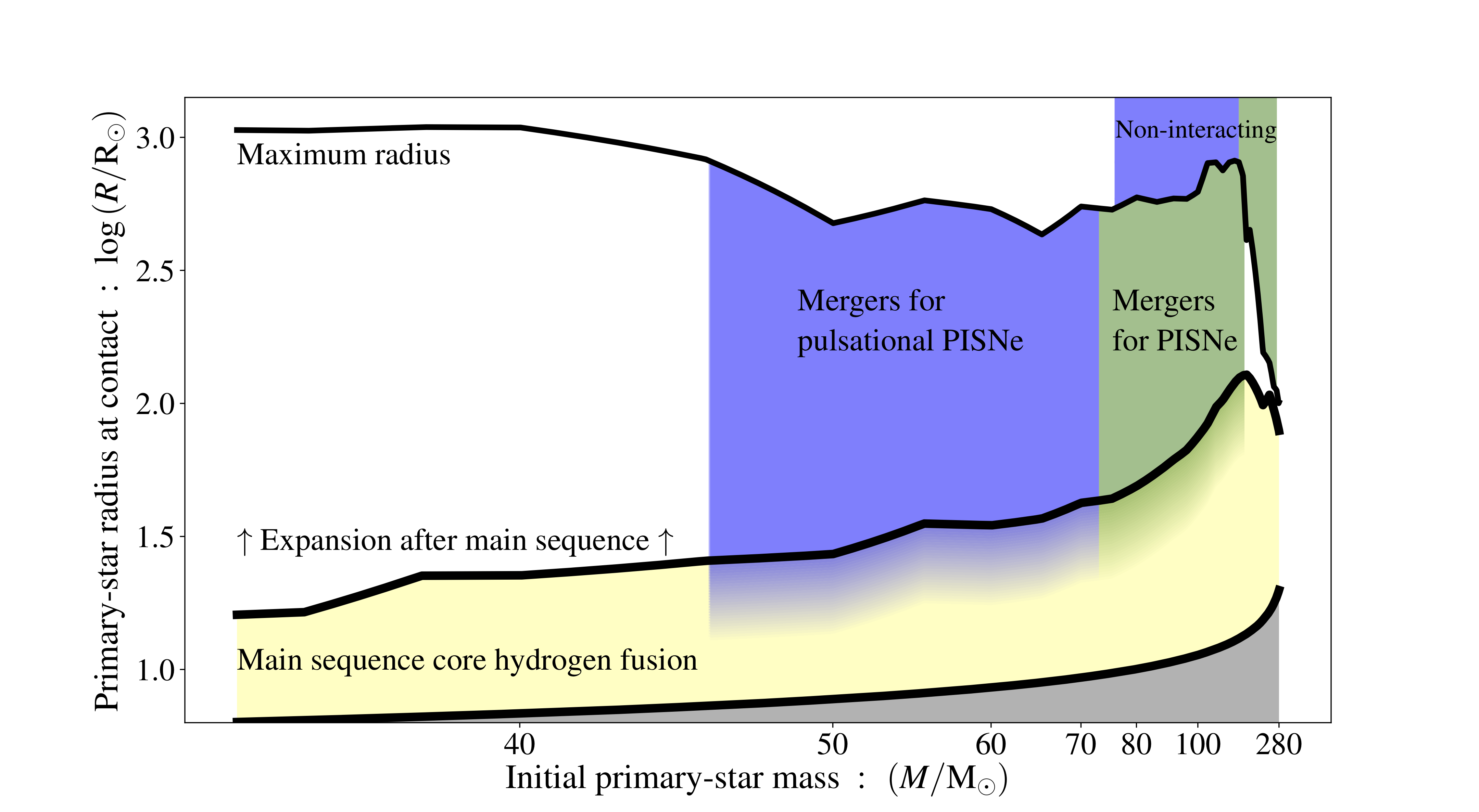} 
\caption{Depiction of the parameter space that is expected to lead to \acp{PISN}, including pulsational \acp{PISN}. The figure shows stellar radii as a function of initial mass. Primary stars evolve vertically upwards in this plot as they expand, and will interact with a companion if they overflow their Roche lobe. The shaded regions indicate where pulsational (blue) and normal (green) \acp{PISN} may be produced. For non-interacting binaries or single stars, this is based on the helium core mass range \citep{Woosley:2017}. For post-main sequence binary mergers the extremes of the shaded regions assume equal-mass mergers. The shading extends to lower radii to indicate a possible contribution from binaries which start mass transfer on the main sequence.  The scalings along both axes are chosen such that equal areas represent equal number density.  Stellar radii and helium core masses for this figure are taken from the MIST library of MESA models  \citep{Choi+2016MIST} for non-rotating models with an initial metallicity of $[\mathrm{Fe/H}]=-0.75$, similar to the SMC metallicity of our MESA models.\label{fig:parameter_space_number}} 
\end{figure*}

Figure \ref{fig:parameter_space_number} illustrates two dimensions of the parameter space for (pulsational) \acp{PISN} arising from non-interacting single stars and merger products.

To investigate the subsequent evolution of such a merger product, we model a case with two identical merging stars, each $60~\rm M_{\odot}$ at \ac{ZAMS}. For a given primary-star mass, an equal-mass case should be the least favourable for retaining hydrogen, since the total fractional core mass at the time of merger will be higher than in cases with non-equal masses, i.e., the fraction of mass in hydrogen at the time of the merger is the lowest. Less massive secondary stars would also retain a larger fraction of their H envelope before the merger because of their reduced stellar winds.

\subsection{Numerical Modeling}

The merger of two stars is complex. Realistic simulations of such events and their outcome are still beyond our reach. However, making reasonable simplifying assumptions, it is possible to study the properties of the products that are the likely outcome of such mergers.  We model a single star and a merger product using Modules for Experiments in Stellar Astrophysics (MESA) \citep{paxton2010modules,paxton2013modules,paxton2015modules,paxton2018modules} (version 10108).   We adopt the same assumptions for the single and merger models, unless stated otherwise.   We use the \texttt{mesa\_67.net} nuclear network, and assume a metallicity of $0.0035$, representative of star-forming regions in the Small Magellanic Cloud, corresponding to $[\mathrm{Fe/H}]\approx -0.76$ ([Z/H] = [Fe/H] adopting solar-scaled abundances as specified in \citealt{Choi+2016MIST}).
For wind mass loss, we use the ``Dutch'' wind scheme in MESA \citep{Glebbeek2009}, reduced by multiplication with a factor of 0.3  \citep{Puls2008,Smith2014ARAA}.   Cool and hot wind schemes are used for effective temperatures below 10,000 K and above 11,000 K, respectively, interpolating when at intermediate temperatures.   For hot winds, the mass loss rate scales with metallicity as $\dot{M}\propto Z^{0.85}$ \citep{Vink2001}.   Convective overshooting extends 0.25 pressure scale heights above convective regions during the \ac{MS} -- specifically until hydrogen is exhausted in the core for the single stars -- and until the moment of merger, which happens shortly after the end of the \ac{MS} in our models.   

\begin{figure*}
\gridline{
			\fig{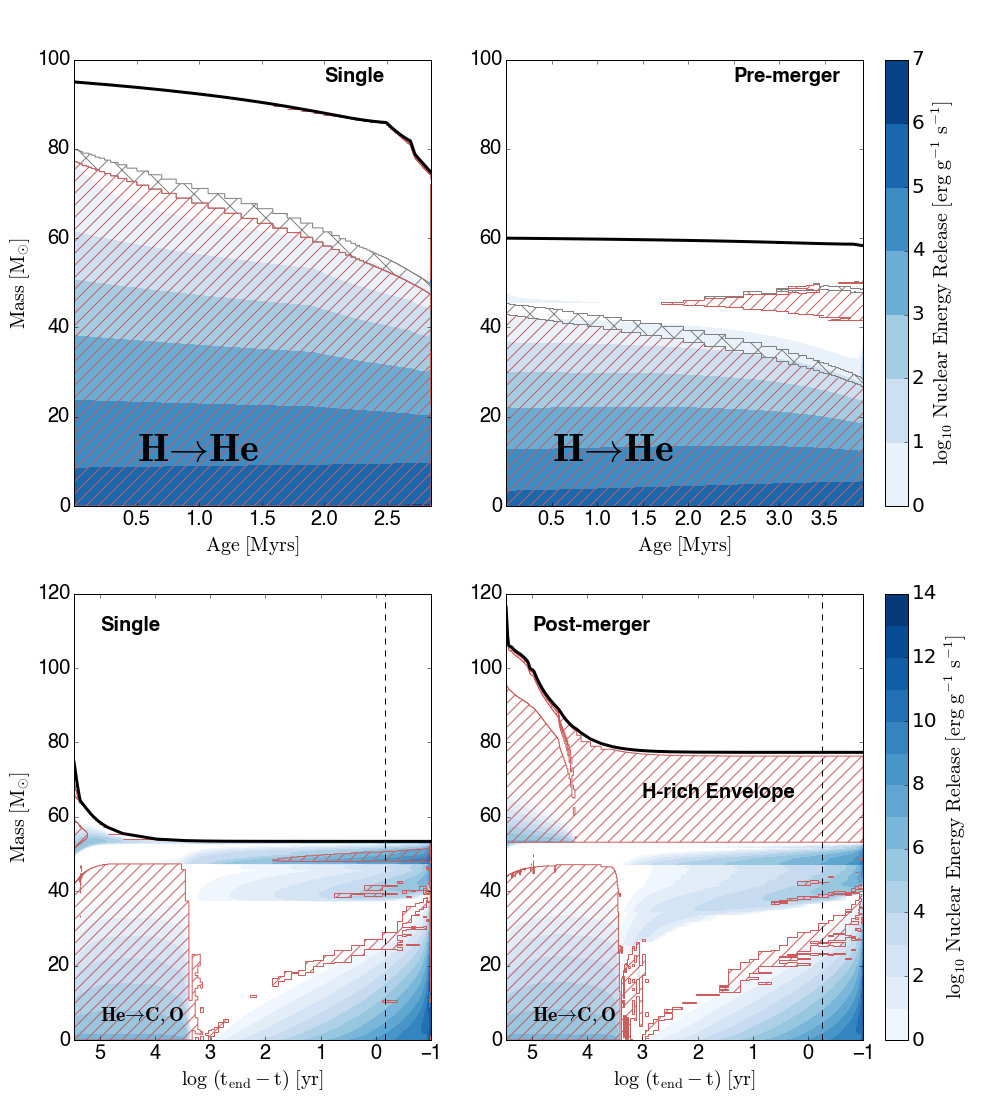}{\textwidth}{}
         }
\caption{Structures of the single (left) and merger (right) models described in the text. Both main sequence (top) and post-main-sequence or post-merger (bottom) evolution are shown. The total mass of each star is shown with a thick solid black curve, nuclear burning regions in shaded blue and convective regions in hatched red. For the post-main-sequence models, the vertical dashed lines (black) show when the central temperature reaches $T_{\rm c}\approx 10^9~\rm\,K$. 
\label{fig:Kips}}
\end{figure*}

The ZAMS mass of our single-star model is $95~\rm M_{\odot}$, and is $60~\rm M_{\odot}$ for each star that we assume to merge.  There is significantly more main-sequence mass loss for the $95~\rm M_{\odot}$ single star than for the $60~\rm M_{\odot}$ pre-merger star (see Figure\ \ref{fig:Kips}), as expected.  

We assume that the structure of the merger product depends on the entropy profile of the merging progenitors, consistent with earlier hydrodynamic situations of mergers between less massive stars \citep{Lombardi2002entropy} and non-rotating massive-star collisions (\citealt{Glebbeek2013}; although see \citealt{Gaburov+2008}). For equal-mass components, this assumption means that the relative internal composition structure of the merger product is initially the same as that of the progenitor star.  Hence we keep the relative internal structure of the pre-merger model fixed whilst doubling the total mass through relaxation, with no abundance change due to nuclear burning during the merger, and with no additional mixing.  However the thermal structure then readjusts in response to the new hydrostatic balance, which can lead to mixing.   Our models are non-rotating and spherically symmetric, but the strong molecular-weight gradients likely suppress rotational mixing \citep[see, e.g.,][and references therein]{justham2014luminous}. We do not remove mass on a dynamical timescale during the merger, as might be expected, but $\approx$7 per cent of the post-merger mass is removed by intense winds during the first $\approx$20 kyr after the merger (just visible in Figure \ref{fig:Kips}), during the time when the model merger product is relaxing towards gravothermal equilibrium. Simulations of head-on mergers between two $40~\rm M_{\odot}$ stars at and just after the end of the main sequence found mass loss of $\approx$8 per cent \citep{Glebbeek2013}.  The amount of mass loss is probably the main uncertainty in our predictions for hydrogen-rich pulsational \acp{PISN}.  This uncertainty affects hydrogen retention, and so whether our merger scenario would produce hydrogen-rich explosions.   

\begin{figure*}
\includegraphics[width=\textwidth]{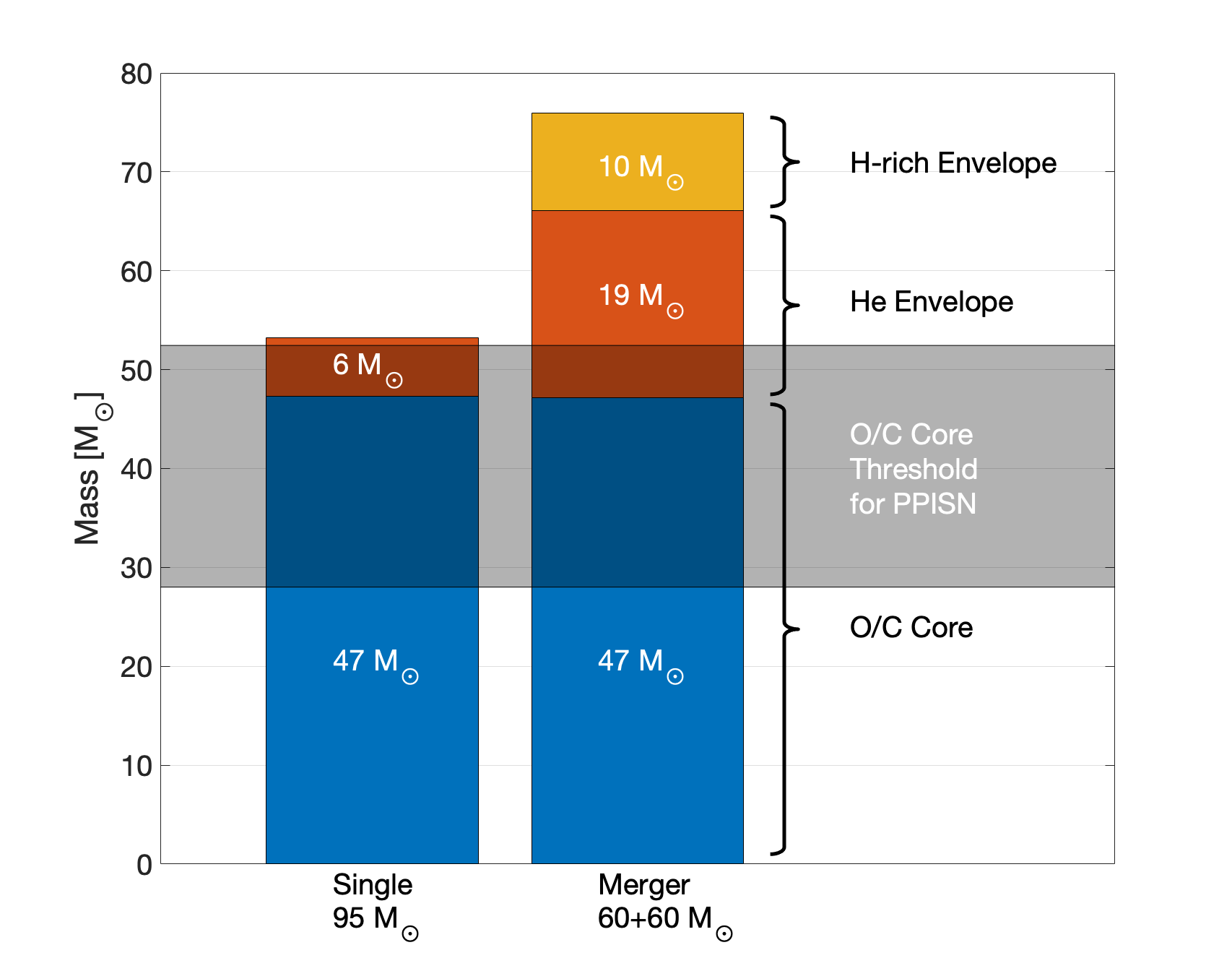}
\caption{Schematic diagram of the dominant compositions of our single (left) and post-merger (right) models at the moment when the central temperature $T_{\rm c}\approx 10^9~\rm K$, less than a year before the first pair-instability driven outburst. The bottom (blue) represents the \ac{O/C} core, whilst the middle (red) and top (yellow) show the masses of helium and hydrogen outside the \ac{O/C} core. The grey shaded region represents the range of masses of \ac{O/C} cores leading to pulsational \acp{PISN} according to \cite{Woosley:2017}: $\lowerLimitPPISN \le M/\rm{M_{\odot}} \le \upperLimitPPISN$. Both cases are expected to lead to a pulsational \ac{PISN}, but the single-star model has less than $0.01~\rm{M_\odot}$ of hydrogen at explosion, while the merger model has approximately $10~\rm{M_\odot}$. }
\label{fig:bars}
\end{figure*}

We model the post-merger evolution with the same assumptions as for the single-star model.  Figure \ref{fig:Kips} shows the stellar structures during these evolutionary phases for both models.  These models were chosen so that the carbon-oxygen core masses are very similar at the onset of pair instability, but the post-merger model retains a massive hydrogen-rich envelope.  The dominant composition structures of these models, just before the onset of the first pulsation, are shown in Figure\ \ref{fig:bars}.  The single-star model has a \ninetyFiveSingleSMCHelium~\Msol\ helium envelope and less than $10^{-2}~\rm M_{\odot}$ of hydrogen, while the post-merger model retains a hydrogen-rich envelope with \sixtySixtyMergerSMCHelium~\Msol\ of helium and \sixtySixtyMergerSMCHydrogen~\Msol\ of hydrogen.

\subsection{Rate estimates} \label{sec:rates}

We estimate the hydrogen-rich pulsational \ac{PISN} rate, $\rate$, as a ratio between the number of pulsational \acp{PISN} from mergers in our scenario, $N_{\rm{PPISN,mergers}}$, and the total number of \acp{CCSN}, $N_{\rm{CCSN}}$, for a fixed amount of star formation:
\begin{equation} \label{eq:rates}
\begin{split}
	\rate & = \frac{N_{\rm{PPISN,mergers}}}{N_{\rm{CCSN}}} \\
		& = f_{\rm binarity}\times f_{\rm primary}\times f_{\rm secondary} \times f_{\rm separation}.
\end{split}
\end{equation}

The factor $f_{\rm{binarity}}$ describes the fraction of massive stellar systems which contain close binaries; we assume it to be $f_{\rm{binarity}}=0.7$ \citep{Sana:2012}. 

The factor $f_{\rm{primary}}$ represents the ratio between the number of binaries in which the initially more massive star is in the correct mass range to produce a pulsational \ac{PISN} if a suitable post-main sequence merger occurs, and the number of stars with the right mass to undergo a \ac{CCSN}.

We assume that stars with \ac{ZAMS} masses in the range $[8,40]~\rm{M_{\odot}}$ undergo \acp{CCSN}, and use the \citet{Salpeter:1955} initial mass function $p(m) \propto m^{-2.3}$.

We estimate $f_{\rm{primary}}$ using terminal-age main sequence helium core masses calculated with MESA. We double those core masses to give notional post-merger core masses, and compare those to the range from \cite{Woosley:2017}. This gives a primary mass range $[\lowerLimitMESA,\upperLimitMESA]~\rm{M_{\odot}}$. As a pessimistic alternative, we only allow the primary to be within $[56,64]~\rm{M_{\odot}}$, with mass ranges symmetric about our calculated example merger model. These assumptions lead to $f_{\rm{primary}} \in [\primaryNarrow, \primaryWide]$.

The factor $f_{\rm{secondary}}$ is the fraction of binaries with a suitable primary in which the lighter companion is sufficiently massive for our merger scenario.  

As a pessimistic assumption we include only mergers between components with nearly equal masses, $q=m_{\rm secondary}/m_{\rm primary} \geq 0.99$ at \ac{ZAMS}. For the more optimistic assumption, we consider that mergers between two stars that have both evolved beyond the main sequence can yield pulsational \ac{PISN} progenitors.  For massive stars, the luminosity-mass relation flattens out and evolutionary timescales vary slowly with mass \citep{Brott2011,Kohler2015}.  This leads to a threshold $q \geq 1+ f_\mathrm{postMS}\,  (\tau/M)\, (dM / d\tau) \approx 0.85$, where $\tau$ is the main sequence lifetime \citep{FarrMandel:2018} and $f_\mathrm{postMS} \approx 0.1$ is the fraction of time the star spends beyond the main sequence. These assumptions lead to $f_{\rm{secondary}} \in [0.01, 0.15]$.

Finally, $f_{\rm{separation}}$ accounts for the fraction of otherwise suitable binary stars with the appropriate separation to merge in the correct evolutionary phase.

We assume the flat-in-the-log distribution of initial separations $p(a) \propto a^{-1}$ \citep[consistent with, e.g.,][]{Sana:2012}.  Depending on whether we require the merger to happen whilst the primary star is close to the end of the \ac{MS} -- within a factor of two in radius -- or optimistically allow for mergers at any point until the star's maximum radial expansion ($\approx 800~\rm{R_{\odot}}$), we find that $f_{\rm{separation}}$ falls in the range $[\separationShort, \separationLong]$.

These assumptions predict a range of rates of suitable mergers leading to pulsational \acp{PISN} that goes from $\rate_{\rm min}=\lowerLimitLogCCSNRate ~ {\rm CCSN^{-1}}$ for conservative assumptions to $\rate_{\rm max}=\upperLimitLogCCSNRate ~ {\rm CCSN^{-1}}$ for more optimistic ones. Empirical estimates indicate that events such as \event~could constitute about $10^{-3}-10^{-2}$ of the Type II \ac{SN} rate \citep{arcavi2017energetic}, where Type II \acp{SN} make up $\approx70$\% of all \acp{CCSN} \citep{Shivvers2017CCSN}. 

Our rate estimate does not include a potentially significant contribution of stellar mergers from binaries with initially closer separations. These massive overcontact binaries \citep{Marchant:2016} may experience several mass transfer episodes during the \ac{MS}, which could lead the stars to approach terminal-age main sequence at the same time. Our rate estimates are also sensitive to mass loss rate prescriptions, which are in turn a function of metallicity. Metallicity-dependent stellar winds reduce both the size of the core and the amount of hydrogen retained in the envelope at the onset of a pulsational \ac{PISN}.  Furthermore, radial expansion, which determines the range of binary separations for which suitable mergers can occur, is also a function of mass and metallicity. We generally expect lower metallicity environments to yield a higher rate of hydrogen-rich pulsational \acp{PISN}.  Their total rate in the local Universe is an integral over all metallicities at which star formation occurs. 

\section{Discussion and Conclusions} 
\label{sec:discussionAndConclusions}

The R139 binary in the Tarantula nebula in the LMC, also known as VFTS 527, consists of two very similar supergiant Of stars of about 60 $M_{\odot}$ and an orbital period of about 150 days 
 \citep[see][]{Taylor+2011,Almeida+2017}.    This is a near-perfect example for our scenario.   Apart from the metallicity, this system is a real-world illustration that systems exist which may follow the merger scenario we have described.   Another well-studied close analogue is WR20a, containing two stars each of $82\ \rm M_{\odot}$  \citep{Rauw+2004,Bonanos+2004,Rauw+2005}, which would be an excellent candidate for a future binary-merger progenitor of a \ac{PISN} if the orbit were slightly wider.  

This merger route is not the only way to potentially increase the rate of pulsational and non-pulsational \acp{PISN}.   Rotational mixing increases the core mass of stars with a given initial mass, which may allow initially less-massive stars to reach the pair-unstable regime, although with less hydrogen in their envelope than for stars of the same mass when evolving without rotational mixing \citep{Langer:2007,deMink:2009,chatzopoulos2012effects,chatzopoulos2012hydrogen,Marchant:2017}.  Runaway mergers of massive stars in stellar clusters have also been discussed as potential progenitors of \acp{PISN} \citep{SPZ+vdH2007,Glebbeek2009}; however, this is not obviously more likely to lead to hydrogen-rich progenitors at the time of the \ac{PISN} than for single stars. 

An intriguing question remains whether iPTF14hls represents a case for a hydrogen-rich PISN resulting from the scenario explored here.  While the metallicity of our models is below the estimated range of $\approx 0.4$ -- $0.9\,Z_\odot$ for the host galaxy of \event\ \citep{arcavi2017energetic}, uncertainties in the metallicity estimate for the progenitor star are significant. We do not claim that this evolutionary scenario has explained all the specific features observed or inferred for the iPTF14hls transient \citep[see, e.g.,][]{arcavi2017energetic,Woosley:2017}, but we consider the possibility worth further modelling and investigation.  Other models, including circumstellar material interaction in a regular \ac{CCSN} and events powered by the spin-down of a magnetar, have not fully explained \event\  \citep{Woosley:2018,Dessart2018}. 

There is tentative observational evidence for an eruption at the location of \event\ 50 years previously \citep{POSSsurvey:1963}. If this was an earlier pulse related to \event, it may be challenging for the simpler pulsational \ac{PISN} to explain it, as much of the hydrogen would likely be expelled during that early pulsation. One very speculative alternative is that this earlier optical transient was related to an extremely late merger, in which case \event\ would be following a very fine-tuned scenario.

Another unusual event that has been discussed as a possible pulsational \ac{PISN} is SN 2009ip \citep{Fraser:2013,Pastorello:2013}, with progenitor metallicity similar to that of the SMC \citep{Pastorello:2013}. It may be interesting to re-consider whether this event, originating from a luminous blue variable star, may also have been a hydrogen-rich pulsational \ac{PISN} from a merger product.

Our scenario does not only produce a pathway for some hydrogen-rich pulsational \acp{PISN}, it also increases the range of initial stellar masses and metallicities from which \acp{PISN} can originate, whether hydrogen-rich or not. \acp{PISN} from mergers may even dominate the \ac{PISN} rate in the local Universe if stellar winds are sufficiently high to suppress \ac{PISN} production at even moderate metallicities in single stars. This formation channel may therefore have significant consequences for the chemical yields from \acp{PISN}. A strong nucleosynthetic signature of enrichment by \acp{PISN} had been expected in low-metallicity stars, but searches for that abundance pattern have had limited success \citep[see, e.g.,][]{NomotoKobayashiTominaga2013}. However, increasing the rate of \acp{PISN} at high metallicity would not exacerbate this problem, because the distinctive \ac{PISN} elemental abundance pattern would be damped when a \ac{PISN} enriches gas that has already been enriched by previous generations of regular supernovae. The new age of wide-field transient surveys is already producing unexpected discoveries, as shown by \event. The ongoing development of such surveys should provide further examples of similar events with which to test our proposal.\\

\acknowledgments
We thank David R. Aguilera-Dena, Ellen Butler, Manos Chatzopoulos, Rob Farmer, Sebastian Gaebel, Ylva G\"{o}tberg, Pablo Marchant, Coenraad Neijssel, Mathieu Renzo and
David Stops for help and discussions, and the reviewer for a constructive and thoughtful report.
We thank the Niels Bohr Institute for its hospitality while part of this work was completed, and the Kavli Foundation and the DNRF for supporting the 2017 Kavli Summer Program.
AVG acknowledges funding support from CONACYT.
SJ and SdM acknowledge funding from the European Union's Horizon 2020 research and innovation programme from the European Research Council (project BinCosmos, Grant agreement No.\ 715063), as well as the Netherlands Organisation for Scientific Research (NWO) as part of the Vidi research program BinWaves (project number 639.042.728).
IM acknowledges partial support from STFC.  This work was partly performed at the Aspen Center for Physics, supported by National Science Foundation grant PHY-1607611, and partially supported by a grant from the Simons Foundation.

\software{\\MESA\citep{paxton2010modules,paxton2013modules,paxton2015modules,paxton2018modules},\\ mKipp  (\doi{10.5281/zenodo.2602098}),\\MESA inlists (\doi{10.5281/zenodo.2644593}).} 




\end{document}